\documentclass[submission,copyright,creativecommons]{eptcs}
\providecommand{\event}{MBT 2013} 
\usepackage{breakurl}             
\usepackage{cite}
\usepackage[english]{babel}

\usepackage{csquotes}

\usepackage[pdftex]{graphicx}

\title{Top-Down and Bottom-Up Approach for\\ Model-Based Testing of Product Lines}
\author{Stephan Wei{\ss}leder
\institute{Berlin, Germany}
\institute{Fraunhofer-Institute FOKUS}
\email{stephan.weissleder@fokus.fraunhofer.de}
\and
Hartmut Lackner 
\institute{Berlin, Germany}
\institute{Fraunhofer-Institute FOKUS}
\email{hartmut.lackner@fokus.fraunhofer.de}
}
\def\titlerunning{Model-Based Testing of Product Lines}
\def\authorrunning{S. Wei{\ss}leder and H. Lackner}
\begin{document}
\maketitle

\begin{abstract}
Systems tend to become more and more complex.
This has a direct impact on system engineering processes.
Two of the most important phases in these processes are requirements engineering and quality assurance.
Two significant complexity drivers located in these phases are the growing number of product variants that have to be integrated into the requirements engineering and the ever growing effort for manual test design.
There are modeling techniques to deal with both complexity drivers like, e.g., feature modeling and model-based test design.
Their combination, however, has been seldom the focus of investigation.
In this paper, we present two approaches to combine feature modeling and model-based testing as an efficient quality assurance technique for product lines.
We present the corresponding difficulties and approaches to overcome them.
All explanations are supported by an example of an online shop product line.
\end{abstract}
 
\section{Introduction}

Today, users of most kinds of products are not satisfied by unique standard solutions, but desire the tailoring of products to their specific needs.
As a consequence, the products have to support different kinds of optional features and, thus, tend to become more and more complex.
At the same time, a high level of quality is expected by the users and has to be guaranteed for all of these product variants.
One example is the German car industry where each car configuration is produced only once on average. 
Summing up, system engineering processes often face challenges that are focused at requirements engineering for product lines and quality assurance, e.g., by testing, at the same time.
This paper deals with the combination of these challenges.
Today, engineering processes are supported by model-driven techniques.
Models are often used to present only essential information, to allow for easy understanding, and to enable formal description and automatic processing.
Models can also be used to describe the features of product lines and the test object as a basis for automatic test design.
Such an approach is also used in this paper.

Product lines (multi-variant systems) are sets of products with similar features, but differences in appearance or price~\cite{CMU_SEI_ProductLines2012}.
There are two important aspects of product lines:
First, users recognize the single product variants as members of the same product line because of their resemblance.
For instance, we recognize cars from a certain manufacturer or certain smart phones although we don't know internal details like, e.g., the power of the engine or the used processors.
Second, the vendors of product lines cannot afford to build every product variant from scratch, but have to strive for reusing components for several product variants.
The product line managers have to try to bring together these two aspects.
For this, they have to know about and manage the variation points of the product line and the relation of variation points and reusable system components.
Feature models can be used to express these variation points and their relations.
They help in making the corresponding engineering process manageable.

Quality assurance is the part of system engineering responsible for ensuring high-quality products, a positive end user experience, and the prevention of damage in safety-critical systems.
Testing is an important aspect of quality assurance.
Since testing is focused on several levels like, e.g., components and their integration, it can be more complex and costly than development.
Because of the afore described growing complexity of systems, it is necessary to reduce the effort for testing without reducing the test quality.
Model-based test design automation is a powerful approach to reach this goal.
It can be used to automatically derive test cases from models.
There are several experience reports to substantiate the success of this technique~\cite{lacknersvacinaweisslederMoTiP10,conformiq_forrester_2010}.

In this paper, we present two approaches to apply automatic model-based test design for the quality assurance of product lines.
All descriptions are supported by an online shop example, i.e., a product line of online shops.
This paper is structured as follows.
In the next section, we describe the example and use it to introduce feature modeling and automatic model-based test design.
In Section~\ref{sec:mbtandmvs}, we present the two approaches for model-based testing of product lines together with an evaluation of their advantages and challenges.
In this paper, we focus on theoretical considerations instead of applying complete tool chains.
Some parts of the projected tool chain, however, can already be used and were applied for our example.
Section~\ref{sec:relatedwork} contains the related work.
In Section~\ref{sec:theend}, we summarize, discuss threats to validity, and present our intended future work.

\section{Fundamentals}
In this section, we define an online shop product line as our running example.
For this example, we assume that we are a provider of online shops.
We offer to install and maintain online shops with different capabilities.
The price depends on the supported set of capabilities.
All of our shops include a catalog that lists all the available products and at least one payment method.
In our example, we allow payment via bank transfer, with ecoins, and by credit card.
The shops can have either a high or a low security level that determine the security of communication.
For instance, using a credit card requires a high security level.
Furthermore, we offer comfort search functions in the shop to support product selection.
Our customers can select a subset of these features for integration into their specific product.

In the following, we use this example to introduce feature models for describing the features, i.e., the variation points of a product line and their relations.
Furthermore, we also use it to introduce state machine models and how to use them for automatic model-based test design.
Finally, we show how to link elements of feature models to elements of other kinds of models.

\begin{figure}[htbp]
\centering
\includegraphics[scale=0.45]{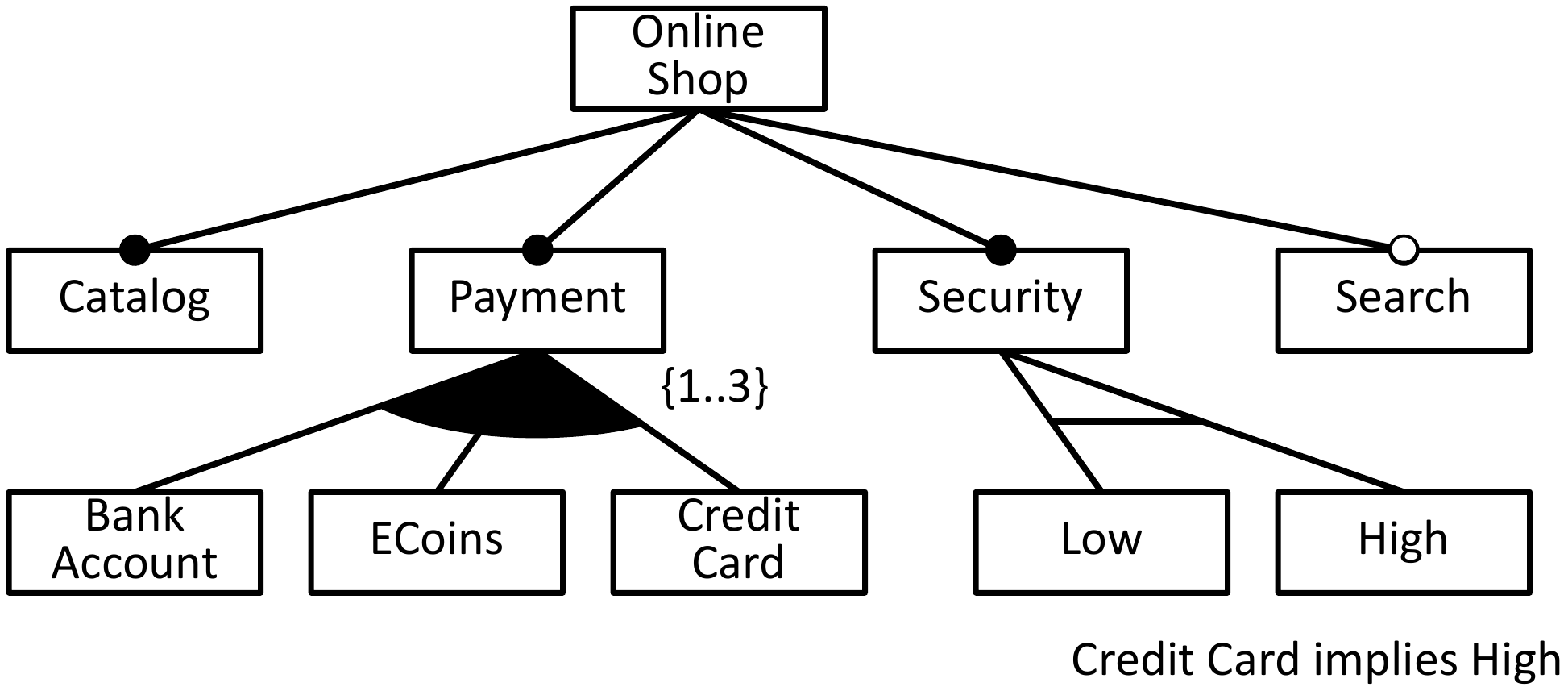}
\caption{Feature model for online shops.}
\label{fig:featuremodel}
\end{figure}

\subsection{Feature Models}
Models are used as directed abstractions of all kinds of artifacts.
The selection of the contained information is only driven by the model's intended use.
Thus, models are often used to reduce complexity and support the user in understanding the described content.
In the following, we present feature models that help in describing all the aforementioned features of our online shop product line.

A feature model is a tree structure of features depicted as rectangles and relations between them depicted as arcs that connect the rectangles.
Figure~\ref{fig:featuremodel} depicts a feature model that contains information about our online shops.
The topmost feature contains the name of the product line.
Four features are connected to it:
The features \emph{Catalog}, \emph{Payment}, and \emph{Security} are connected to the top-most feature by arcs with filled circles at their end, which describe that these three features are mandatory, i.e., exist in every product variant.
The \emph{Search} feature is optional, which is depicted by using an arc that ends with an empty circle.
This hierarchy of features is continued.
For instance, the feature \emph{Payment} contains the three subfeatures \emph{Bank Account}, \emph{ECoins}, and \emph{Credit Card}, from which at least one has to be selected for each product variant.
The subfeatures \emph{High} and \emph{Low} of the feature \emph{Security} are alternative, which means that exactly one of them has to be chosen for each product variant.
Furthermore, there is a textual condition that states that credit cards can only be selected if the provided security level is high.

Summing up, feature models are a simple way to describe the variation points of a product line and their relations at an abstract level that is easy to understand.
Their semantics, however, only consist of rectangles and arcs with no links to system engineering-relevant aspects such as requirements or architecture models.
The importance of feature models for the system engineering process only becomes real if they are integrated into the existing tool chain.
This integration is done by linking the features of the feature model to other artifacts like, e.g., requirements in DOORS~\cite{toolRationalDOORS}.
There also exists corresponding tool support~\cite{toolPureVariants2012,toolFeatureMapper2012}.
In our approach, we link features to elements of state machines of the Unified Modeling Language (UML)~\cite{uml24} to steer automatic model-based test design for product lines.

\begin{figure}[htbp]
\centering
\includegraphics[scale=0.45]{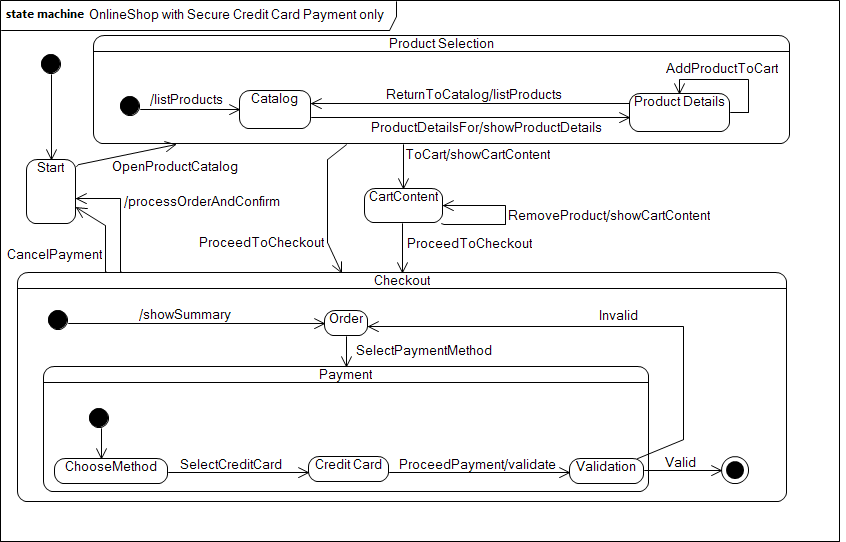}
\caption{Online shop state machine diagram for one product variant.}
\label{fig:eShopSDM}
\end{figure}

\subsection{Automatic Model-Based Test Design}
Models can also be used for testing.
The corresponding technique is called model-based testing (MBT) and there are many different approaches to it~\cite{Utting_2012_taxonomy}.
Several kinds of models are applicable for MBT like, e.g. system models or environment models~\cite{weisslederlacknerMOTES10}.
Furthermore, different modeling paradigms are applicable like, e.g., state charts, petri nets, or classification trees.

For our online shop system, we focus on the automatic derivation of test cases based on structural coverage criteria that are applied to state machines.
Figure~\ref{fig:eShopSDM} shows such a state machine.
The behavior depicted in this state machine corresponds to one product variant of our online shop product line that only allows to pay per credit card and does not include the search function:
A user of the online shop can open the product catalog (\emph{OpenProductCatalog}).
In this state, the user can select products and have a look at their details (\emph{ProductDetailsFor}). 
In the detail view, the user can decide to add the product to his shopping cart (\emph{AddProductToCart}).
After the user selected products, he can decide to remove some selected elements again (\emph{ToCart}, \emph{RemoveProduct}) or to finish the transaction (\emph{ProceedToCheckOut}).
For paying, the user first has to select a payment method (\emph{SelectPaymentMethod}).
For the depicted shop variant, the user is only allowed to select the credit card payment method (\emph{SelectCreditCard}).
Afterwards, the system validates the entered user data and if they are valid (\emph{Valid}), then the order is processed and a confirmation message is shown to the user.
Finally, the user is forwarded to the initial page of the shop.
Like depicted in the state machine, the user has the option to cancel the process and return to the start page during the checkout process.

This model can be used for automatic test design.
As stated above, there are several ways to do so.
A widely used approach is to apply coverage criteria like, e.g., All-Transitions~\cite{1200168} to the state machine.
A test generator then tries to create paths on the state machine that cover all transitions of the state machine.
These paths can be executed by using the sequence of triggers that correspond to the path transition sequence.
Afterwards, the created paths are translated into test cases of the desired target language that can be used, e.g., for documentation or test execution.
There are several automatic model-based test generators available like, e.g., the Conformiq Designer~\cite{toolTIConformiq_4_4} or ParTeG~\cite{toolParteg}.

Automatic model-based test design for single product variants is well-known.
In this paper, we do focus on how to use this technique for product lines.

\subsection{Linking Feature Models and Models for Test Generation}
\begin{figure}[htbp]
\centering
\includegraphics[scale=0.5]{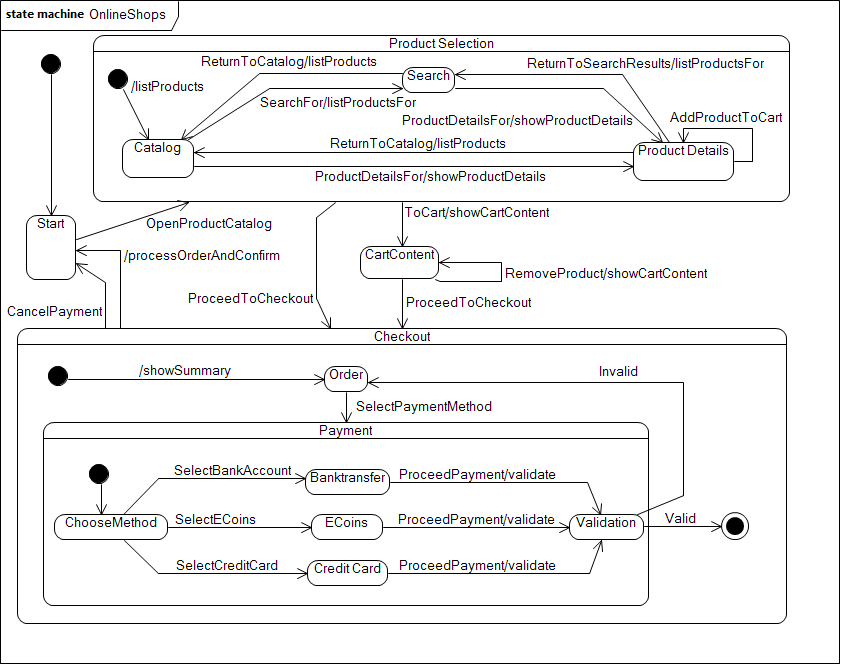}
\caption{Online shop 150\% state machine.}
\label{fig:eShop}
\end{figure}

In order to apply model-based test generation to product lines, the model for test generation has to be linked to the feature model.
One straight forward approach is to also describe all variation points in the state machine, i.e., the possible behavior of the system under test, and to link the features of the feature model to these variation points.
Figure~\ref{fig:eShop} depicts a model that contains the behavior of all product variants.
Because more than one variant is described, such models are called 150\% models.
As one can see, the depicted state machine contains elements that correspond to the aforementioned variation points of the online shop product line.
However, it is not a complete model of our system as it lacks the information from the feature model like, e.g., the relations of the features and the corresponding information about the validity of feature selections.
To resolve this, we connect the features of the feature model to the 150\% state machine with logical expressions.

\begin{figure}[htbp]
\centering
\includegraphics[scale=0.6]{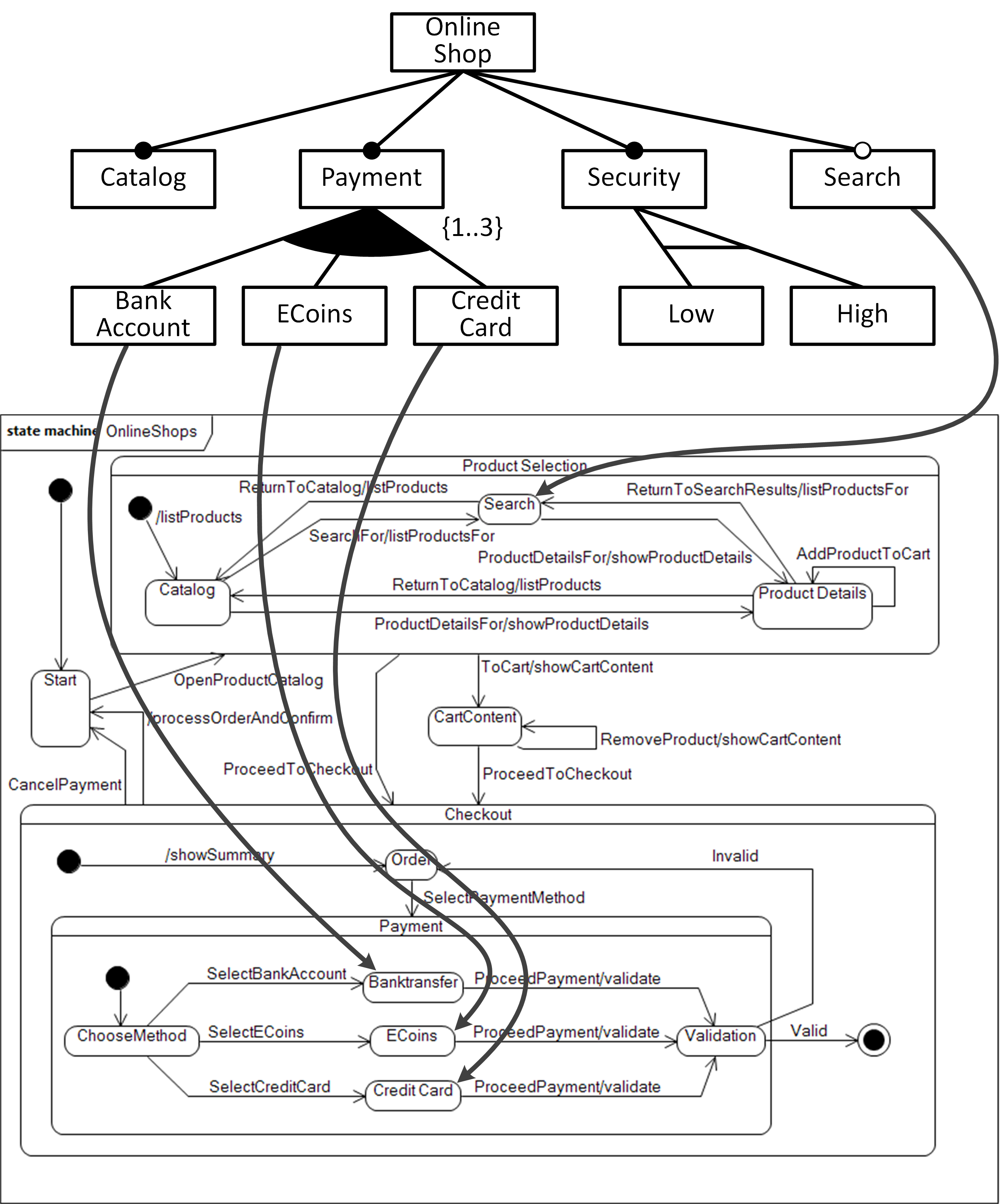}
\caption{Mapping the feature model to the 150\% state machine.}
\label{fig:mapping}
\end{figure}

Mapping features to other model elements can require complex logical expressions and, thus, can become complex.
For reasons of simplicity, we link the models by a mapping model that links features of the feature model to one or more model elements of the 150\% model.
The application of a configuration to the state machine results in a 100\% model by deleting all model elements that are not associated to that particular configuration.
Figure~\ref{fig:mapping} depicts the mapping of the feature model to the 150\% model.
Using this mapping, it is possible to select valid variants or sets of them, to derive corresponding 100\% state machines, and to automatically derive test cases from them.
As described in the related work in Section~\ref{sec:relatedwork}, there are already some approaches that head into the same direction.
For instance, the product-by-product approach creates all valid product variants, derives the corresponding 100\% models, and applies model-based test generation to each model.
However, this approach corresponds to a brute force approach, which is infeasible for larger systems.
There are several approaches of how to design test cases for such linked models.
In the following, we present and compare two more mature test design approaches for product lines.

\newpage

\section{Applying Model-Based Test Design to Product Lines}
\label{sec:mbtandmvs}
In the previous section, we described how to use feature models to describe variation points of product lines, how to use state machines for automatically designing test cases, and how to link feature models and state machines.
By this, we provided the infrastructure for automatic model-based test design for product lines.
There are, however, several processes of actually deriving test cases from the combination of these two kinds of models.
In the following, we are going to present two different approaches and to evaluate their pros and cons using the described example.

\subsection{Top-Down Approach}
\begin{figure}[htbp]
\centering
\includegraphics[scale=0.7]{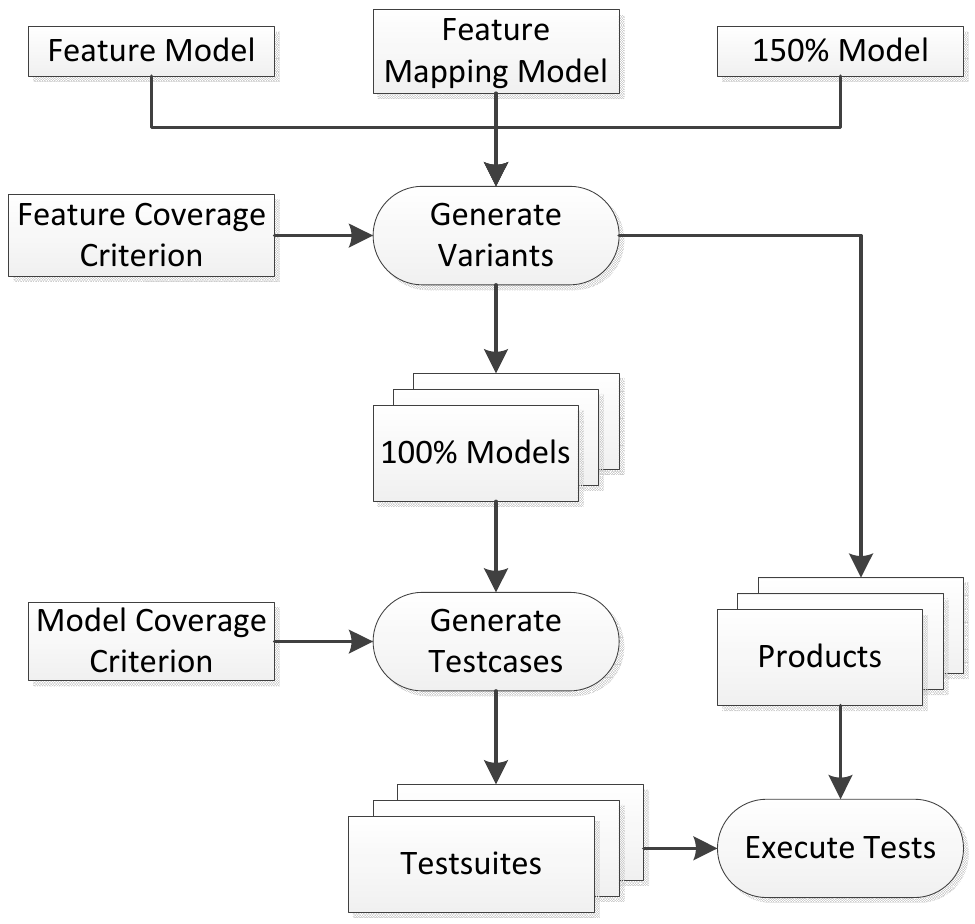}
\caption{Top-down approach for test generation.}
\label{fig:TopDown}
\end{figure}

In the top-down approach, we first derive a set of product variants from the feature model, derive the set of corresponding  100\% models, and apply standard model-based testing to each 100\% model.
Automatic model-based test generation is often driven by applying coverage criteria to models.
This approach as presented for state machines can also be applied to feature models.
The coverage criteria are used to measure to which degree the product variants represent the product line, i.e., the set of all possible product variants.
They can also be used to automatically derive a representative set of product variants~\cite{DBLP:conf/vamos/OsterZML11}.
Using the links between feature model and 150\% state machine allows for automatic derivation of the corresponding 100\% state machines for each generated product variant.
For each of these 100\% state machines, the presented standard approach of test generation based on structural coverage criteria can be applied.
Afterwards, the generated test suites can be executed for the corresponding product variants.
Figure \ref{fig:TopDown} depicts this approach.

For evaluating the strengths and weaknesses of the top-down approach, we consider two aspects:
\textbf{a)} To which degree and with which efficiency are product variants covered?
\textbf{b)} To which degree and with which efficiency is the system behavior covered? 
For \textbf{a}, the coverage criteria applied to the feature model directly determine the coverage of the feature model.
The answer to \textbf{b} additionally depends on the relative strength of the coverage criterion that is applied to each 100\% model.
Furthermore, there will be overlap between the behavior of the variants, which is going to be tested twice.
So the presumption is that the generated test cases will not be very efficient, i.e., several parts will be tested twice without additional gain of information.

\begin{figure}[htbp]
  \centering
  \includegraphics[scale=0.5]{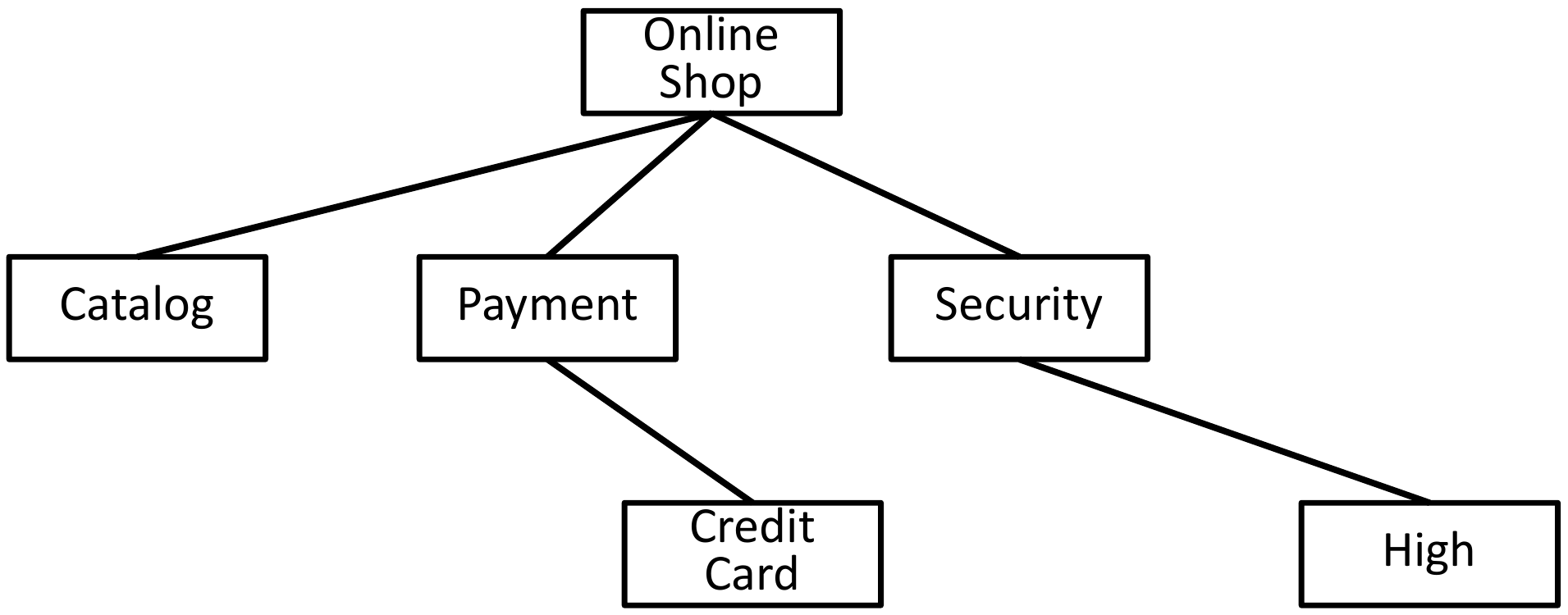} 
  \caption{Product variant (i).}
  \label{fig:varianti}
\end{figure}

\begin{figure}[htbp]
  \centering
  \includegraphics[scale=0.5]{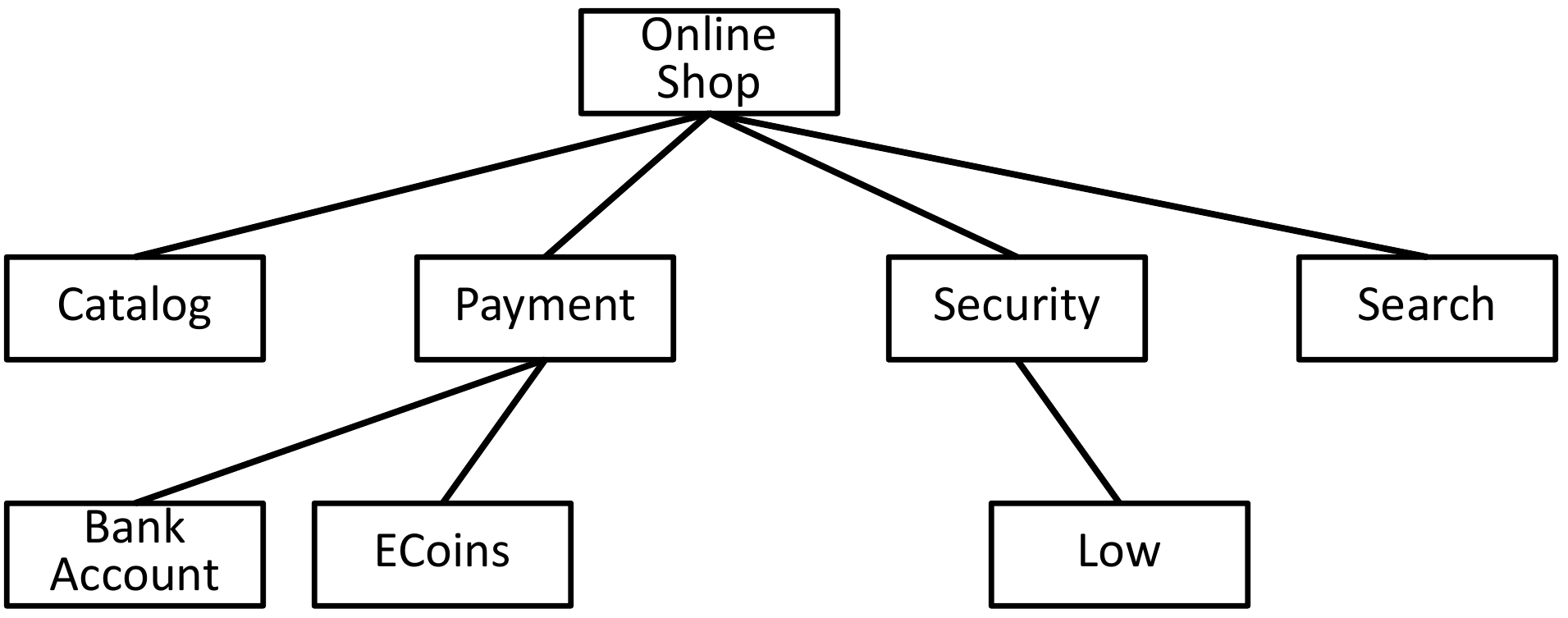}  
  \caption{Product variant (ii).}
  \label{fig:variantii}
\end{figure}

For our example of the online shop product line, we run test generation with the following combination of coverage criteria:
We apply the two coverage criteria \emph{All-Features-Selected} and \emph{All-Features-Unselected} to the feature model to derive variants.
Then, we derive the corresponding 100\% state machines using the mapping from the feature model to the 150\% state machine and generate tests using the coverage criterion \emph{All-Transitions}~\cite{1200168} on every generated 100\% state machine.
To the best of our knowledge, the research on coverage criteria on feature models is still in an immature state and, thus, references to such coverage criteria are rare~\cite{DBLP:journals/sqj/LochauOGS12}.
In contrast to existing work on coverage criteria, it is also important to focus on covering features by not selecting them.
The mentioned two coverage criteria are correspondingly focused on selecting and not selecting all features of a feature model, respectively.
The two coverage criteria \emph{All-Features-Selected} and \emph{All-Features-Unselected} can be satisfied by two product variants in which the following optional features are selected: 
\textbf{(i)} Credit Card Payment and High Security; \textbf{(ii)} Bank Transfer, ECoins, Low Security, and Search.
Figures~\ref{fig:varianti} and~\ref{fig:variantii} show the corresponding variant models.
\\
For each variant, one test case is enough to cover all transitions of the corresponding 100\% state machine.

For \textbf{(i)}, the sequence of events to trigger the test case is as follows (see 100\% model in Figure~\ref{fig:eShopSDM}):\\
\emph{OpenProductCatalog; ProductDetailsFor; AddProductToCart; AddProductToCart; ReturnToCatalog; ToCart; RemoveProduct; ProceedToCheckout; CancelPayment; OpenProductCatlog; ProceedToCheckout; SelectPaymentMethod; SelectCreditCard; ProceedPayment; Invalid; SelectPaymentMethod; SelectCreditCard; ProceedPayment; Valid}.

For \textbf{(ii)}, the sequence is as follows (no corresponding 100\% model depicted - please refer to the corresponding parts of the 150\% model in Figure~\ref{fig:eShop}):\\
\emph{OpenProductCatalog; SearchFor; ProductDetailsFor; AddProductToCart; AddProductToCart; ReturnToSearch; ReturnToCatalog; ToCart; RemoveProduct; ProceedToCheckout; CancelPayment; OpenProductCatalog; ProductDetailsFor; AddProductToCart; ReturnToCatalog; ProceedToCheckout; SelectPaymentMethod; SelectBankAccount; ProceedPayment; Invalid, SelectPaymentMethod; SelectECoins; ProceedPayment; Valid}.

All features have been selected as well as deselected.
Both test cases together cover all 22 explicitly triggered transitions of the 150\% model.
They have a total length of (i:19 + ii:24) 43 event calls, which is almost twice the size of the lower boundary. 
Since we created two product variants for testing instead of the 20 possible ones, however, this approach is still far more efficient than the brute force approach.

Our tool chain that supports the described test generation approach is currently under construction.
However, we already manually created the two 100\% state machines and used the Conformiq Designer~~\cite{toolTIConformiq_4_4} to automatically design tests.
The used coverage criterion is All-Transitions.
For variant \textbf{i}, the test generator created seven test cases that comprise altogether 28 event calls. 
In case of variant \textbf{ii}, eleven test cases with 42 event calls were generated. 
Since the Conformiq Designer is not tailored to find as many test steps per test case as possible, this deviation from our theoretical considerations are no surprise.
After all, all transitions were covered for both cases.

\subsection{Bottom-Up Approach}
The idea of the bottom-up approach is contrary to the top-down approach.
Here, we create test cases based on the 150\% state machine and match the resulting sequences to single product variants, afterwards.
The idea is simple, but the generated paths of the state machine cover elements that are linked to different features and the state machine does not provide means to check if all these features can be combined in one valid product variant.
As a result, one has to include the conditions that are expressed in the feature model into the 150\% state machine.
This is done by expressing the selection and deselection of a feature to a variable with the value $1$ and $0$, respectively.
\begin{figure}[htbp]
\centering
\includegraphics[scale=0.7]{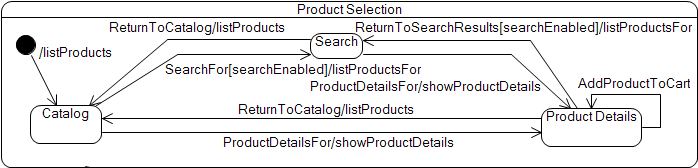}
\caption{Enriched part of the 150\% state machine for automatic test generation of only valid product variants.}
\label{fig:FeatureMerge}
\end{figure}
Figure~\ref{fig:FeatureMerge} depicts such an enrichment on an excerpt of the 150\% state machine.
The shown composite state was enriched with information from the feature model by adding a guard to all transitions leading to the state \emph{Search}, which corresponds to the search feature. 
Now, the tests cover the search function only if the corresponding guard is set to true.
Setting the guard variable to a value is possible only once at the beginning of the state machine.
As a consequence, the variable and the feature selection will be consistent for the whole test case.
Relations between features can also be expressed in the guard, e.g., by stating that the value of the variable corresponding to an alternative feature has a different value.

\begin{figure}[htbp]
\centering
\includegraphics[scale=0.7]{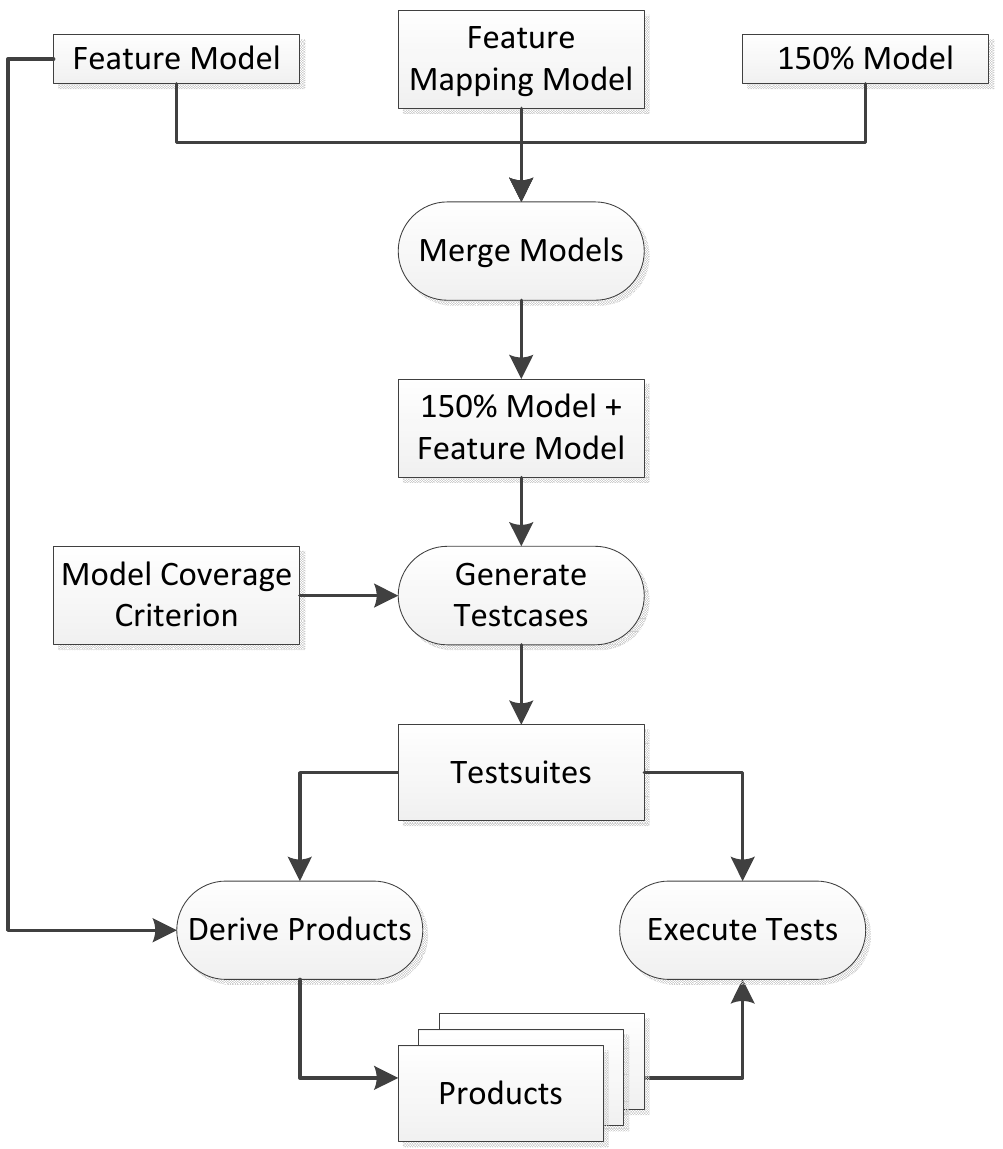}
\caption{Bottom-up approach for test generation.}
\label{fig:BottomUp}
\end{figure}

This enables the generator to choose from any valid configuration for finding a new test case.
Since we did not generate the product variants from the feature model, we had to retrieve the necessary product variants for test execution from the test cases.
Conformiq supports this task because all initial variable values can be stored into the prolog of a test case.
Figure~\ref{fig:BottomUp} shows the workflow for the bottom-up approach.

In the following, we present the input sequence for a test case that covers all transitions in the enriched 150\% model:\\
\emph{OpenProductCatalog, ProductDetailsFor, ReturnToCatalog,
SearchFor, ProductDetailsFor, AddProductToCart, AddProductToCart, ReturnToSearchResults, 
ReturnToCatalog, ToCart, RemoveProduct, ProceedToCheckout, 
SelectPaymentMethod, SelectBankAccout, ProceedPayment, Invalid, 
SelectPaymentMethod, SelectECoins, ProceedPayment, Invalid,
CancelPayment, OpenProductCatalog, Proceed\-To\-Checkout, 
SelectPaymentMethod, SelectCreditCard, ProceedPayment, Valid}.\\
This test case has 27 test steps, covers all 22 event calls of the 150\% model, and requires only one product variant for test execution in which all features except \emph{Low} (Security) have been selected. 

Again, we use the Conformiq Designer for test generation focused on covering all transitions.
The result of the test generation are twelve test cases with overall 59 event calls.

\subsection{Comparison}
Here, we summarize the first evaluations of both approaches and compare them to each other.

Concerning our theoretical considerations and the manually created test cases, the top-down approach results in two test cases that use 43 event calls and the bottom-up approach results in one test case with 27 event calls.
From the perspective of redundancy, the bottom-up approach seems to be more efficient than the top-down approach.
On the one hand, this always depends on the used coverage criteria for the feature model.
For instance, a weaker coverage criterion that is satisfied by only one product variant, can lead to a more efficient result for the top-down approach.
On the other hand, this is no generally applicable solution because the importance of single variants for the behavior is not easy to determine.
The bottom-up approach abstracts from this issue because one does not have to define the coverage criterion on the feature model in the first place, but it does not necessarily cover all features of the feature model.
As a result, it seems that the personal notion and the importance of the covered aspects is important:
If feature coverage is important, the top-down approach is more suitable.
If efficiency and covered behavior is more important, the bottom-up approach is more suitable.

The application of the Conformiq Designer shows partly similar though different results.
For the two 100\% models, 18 test cases with 70 event calls were generated.
For the 150\% model, twelve test cases with 59 event calls were generated.
The main reason for the deviation to the manually created test cases is the breadth-first search approach of Conformiq and our approach of finding the minimal number of test cases.
Furthermore, the Conformiq Designer created test cases for two product variants for the bottom-up approach.
Interestingly, both variants include high security and differ only in the selection of the feature \emph{Credit Card}.
This distinction is unnecessary and would have been avoided by a human designer.
This issue leaves room for future improvements.

\section{Related Work}
\label{sec:relatedwork}
In this section, we present the related work.
We present standard approaches to model-based testing, cite work about feature modeling, and name approaches to combining both.

Testing is one of the most important quality assurance techniques in industry.
Since testing often consumes a high percentage of project budget, there are approaches to automate repeating activities like, e.g., regression tests.
Some of these approaches are data-driven testing, keyword-driven testing, and model-based testing.
There are many books that provide surveys of conventional standard testing~\cite{1355340,338330,539883} and model-based testing~\cite{broy_05_model-based,1200168,zanderschieferdeckermosterman2011}.
In this paper, we use model-based testing techniques and apply them to product lines.
Modeling languages like the Unified Modeling Language (UML)~\cite{uml24} have been often used to create test models for testing.
For instance, Abdurazik and Offutt~\cite{offutt99generating} automatically generate test cases from state machines.
We also apply state machines of the UML.

Feature models are commonly used to describe the variation points in product lines.
There are several approaches to apply feature models in quality assurance.
For instance, Olimpiew and Gomaa~\cite{1083279} deal with test generation from product lines and sequence diagrams.
In contrast to that, we focus on UML state machines and describe different approaches for combining both.
In contrast to sequence diagrams, state machines are commonly used to describe a higher number of possible behaviors, which make the combination with feature models more complex than combining feature models and sequence diagrams.
As another example, McGregor~\cite{McG01c} shows the importance of a well-defined testing software product line process.
Just like McGregor, the focus of our paper is only investigating the process of actually creating tests rather than defining the structural possible relations of feature models and state machines.
Pohl and Metz\-ger~\cite{1183271} emphasize the preservation of var\-i\-ability in test artifacts of software product line testing.
As we derive test case design from models automatically, this var\-i\-ability is preserved.
Lochau et al.~\cite{DBLP:journals/sqj/LochauOGS12} also focus on test design with feature models.
In contrast to our work, they focus on defining and evaluating coverage criteria that can be applied to feature models.
In the presented top-down approach, we strive for using such coverage criteria on feature models for the automation of test design.
Cichos et al.~\cite{completeSPLTestSuite2011} also worked on an approach similar to the presented bottom-up approach.
Their approach, however, requires that the used test generator has to be provided a set of product variants to derive 100\% models from the 150\% model for automatic test generation.
As a consequence, the test generator requires an additional input parameter and (as the authors state) no standard test generator can be applied for their approach.
In contrast, both of our approaches allow for integrating commercial off-the-shelf test generators like in our case, Conformiq~\cite{toolTIConformiq_4_4}.
One of the most important aspects in our work is the ability to integrate our approach into existing tool chains.
In~\cite{weissleder_sokenou_schlingloff_MoTiP2008}, we already addressed model-based test generation for product lines.
However, back then we focused on reusing state machines in multi-variant environments instead of describing the different automatic test design approaches for product lines.

\section{Summary, Discussion, and Future Work}
\label{sec:theend}

In this paper, we presented different approaches to the automatic test design for product lines.
We described the state of the art, presented the general idea of linking feature models to other system artifacts, and presented two approaches to use this linking for automatic test design.
Our main contributions are the definition and comparison of the presented approaches using a small example.

The presented outcomes are theoretical considerations and first test generation results using the Conformiq Designer.
Some steps of the proposed tool chains are still under construction and the corresponding intermediate results were, thus, partly created manually.
If the missing parts of the tool chains will be developed, we will be able to run larger case studies for the comparison of both approaches automatically.
Furthermore, there are more approaches than the presented ones of automatically designing tests for product lines that were not evaluated here.
To name just one example, single steps in our automatic tool chain could also be replaced by manual experience-based steps.
Another point to discuss is the degree of reusability in the source code.
As mentioned in the beginning of this paper, reusing system components is an important aspect in managing product variants engineering for product lines.
If the components, however, were not reused adequately and copy\&paste was applied, instead, then covering the behavior at the source code level for one product variant does not necessarily imply covering the very same behavior at the source code level for another product variant.
A solution to this issue would be to also integrate the relations from features in the feature model to variation points in the source code.

In the near future, we plan to finish the development of both proposed tool chains.
The tool chains are intended to provide the glue between existing tools for feature modeling like, e.g., pure::variants~\cite{toolPureVariants2012} or the FeatureMapper~\cite{toolFeatureMapper2012}, and automatic test generators like, e.g., the Conformiq Designer~\cite{toolTIConformiq_4_4} or ParTeG~\cite{toolParteg}.
Besides the comparison of the two approaches, the single approaches contain enough room for further investigations.
For instance, one interesting question for the top-down approach is if it is advisable to apply strong coverage criteria on the feature model and weak ones on the 100\% models or vice versa.
For the bottom-up approach, an interesting task is to retrieve a minimal number of product variants from test cases generated from the 150\% model.
As stated above, we also plan to run further experiments to evaluate the pros and cons of the presented approaches.

\bibliographystyle{eptcs}
\bibliography{biblio}
\end{document}